\documentclass[11pt, a4paper, onecolumn, oneside]{article} 

\usepackage[scale=0.775]{geometry}
\usepackage[labelfont=bf,textfont=md]{caption}
\usepackage{url,parskip} 
\usepackage[usenames,dvipsnames]{xcolor} 

\usepackage{cite}
\usepackage{graphicx,epsfig,epstopdf}
\usepackage{amssymb,amsmath}
\usepackage{multicol}
\usepackage{subfigure}
\usepackage[a4paper=true,pagebackref=false]{hyperref} 
\definecolor{linkcolour1}{rgb}{0.2,0.5,0.2} 
\definecolor{linkcolour2}{rgb}{0.2,0.6,0.2} 
\hypersetup{colorlinks,breaklinks,urlcolor=linkcolour1,linkcolor=linkcolour2,anchorcolor = green, citecolor = blue, filecolor = red} 

\title{Dynamical systems of modified Gauss-Bonnet gravity: cosmological implications}

\author{Ashutosh Singh${}^{1}$\footnote{ashuverse@gmail.com}
\\
${}^{1}$Centre for Cosmology, Astrophysics and Space Science,\\ GLA University, Mathura, Uttar Pradesh 281406, India \\
}

\date{}
\begin{document}
\maketitle

\begin{abstract}
In this paper, we derive the field equations of modified Gauss-Bonnet gravity termed as $f(R,G)$ gravity for the non-flat Friedmann-Robertson-Walker (FRW) spacetime. We utilize the dynamical system approach to study the cosmic dynamics of two different class of $f(R,G)$ models composed of radiation and matter (cold dark matter and baryonic matter). The linear perturbations around the fixed points are studied to explore the corresponding stability of points. The cosmological implications are studied in $f(R,G)=f_0R^nG^{1-n}$ and $f(R,G)=f_0R^\alpha+f_1G^\beta$ models to identify the qualitative evolution of universe with the flat-FRW spacetime. The qualitative differences between the considered class of models are discussed in detail. The fixed points corresponding to the late-time accelerated and radiation phase of the universe will exist in the model but, the existence of fixed point corresponding to the matter dominated phase will depend on the functional form of $f(R,G)$. Furthermore, the autonomous systems are utilized to study the cosmographic parameters along with the statefinder diagnostic.  
\end{abstract}

\section{Introduction}
\label{sec1}
The high precision astronomical findings highlight that the universe expansion is accelerating \cite{Riess+1998,Perl+1999,netal}. In the General relativity framework, this kind of expansion may be explained by the introduction of an extra degree of freedom such as a new field or fluid giving rise to negative pressure necessary for the accelerated expansion \cite{Bamba+2012,Capozziello+2019}. This field or fluid has been termed as the dark energy \cite{noo2017}. Among different proposals, the model with positive cosmological constant fits well with the available observational data \cite{pje,smc,a}. However, this cosmological constant model (also known as $\Lambda$ cold dark matter model) is affected by the fine-tuning problem and cosmic coincidence problem \cite{sw}. The fine-tuning problem originated due to the difference between the vacuum energy density predicted from the theories in particle physics with its observed value \cite{sw,lrr2001}. The coinciding values of the energy densities of vacuum  and matter at the present time in the cosmic history results into the cosmic coincidence problem \cite{sw,lrr2001,jgav}. These problems of the $\Lambda$ cold dark matter ($\Lambda$CDM) model motivates to search for the alternative paradigms in the modern cosmology. \\
In the alternative theoretical scenarios, modifications of the General relativity at the cosmologically relevant scales may provide the mechanism for cosmic accelerated expansion of the universe \cite{revm,lrr,ds3,nps,a2,adf,fg1,fg2,fg4,prd83,fgd,scjpm,mgg,sdat,epjc22,prd87,bbepjc,arxjls,epjc77,st23,sray,cs1,cs2,a11,ll1,ll2,bf1,bf2,ee1,ee2,3,bm,1,2,48,49,ext6,dcm2,a5a,a6,a10a,dcm1,a15,a16,adixit2,a12,adixit1,arle24}. The $f(R)$ theory \cite{revm,lrr,ds3,nps,a2} is the simplest modified gravity theory obtained by replacing the Ricci scalar $(R)$ by $f(R)$ in the Einstein-Hilbert action of the General relativity. Among other extensions of the General relativity, the action may involve second order invariant involving the Ricci scalar, Ricci tensor $(R_{ij})$ and the Riemann tensor $(R_{ijkl})$ \cite{adf}. The appearance of Gauss-Bonnet invariant in the gravitational action may lead to a modified theory which could be of interest to address different issues of the General relativity at different physical scales \cite{fg1,fg2,fg4,prd83,fgd,scjpm,mgg,sdat,epjc22,prd87,bbepjc,arxjls,epjc77,st23,sray}. The Gauss-Bonnet invariant $(G)$ defined as $G\equiv R^2-4R_{ij}R^{ij}+R_{ijkl}R^{ijkl}$ may arise naturally in the gauge theories like Chern-Simons \cite{cs1,cs2,a11}, Lovelock \cite{ll1,ll2} or Born-Infeld gravity \cite{bf1,bf2}. The topological nature of this term allows to simplifying the dynamics by reducing the order of  equations of gravitational motion. In $f(R,G)$ theory, the presence of both Ricci scalar and Gauss-Bonnet invariant term may lead to the non-trivial dynamics in the models \cite{ee1,ee2,3,bm,1,2}. The cosmological properties of modified Gauss-Bonnet gravity model, in principle, explains the primordial inflationary phase and the accelerated universe expansion of late-times \cite{ee2,48,49}. Cruz-Dombriz and Saez-Gomez \cite{Cqg29} analyzed the homogenous and isotropic perturbations of the Hubble parameter and matter density around the power-law and de Sitter solutions in context of $f(R,G)$ gravity. The power-law solutions and the corresponding complexity of perturbed equation coefficients may lead to $f(R,G)$ forms such as $f(R,G)=f_0R^\alpha G^\beta$ and $f(R,G)=f_1(R)+f_2(G)$ \cite{Cqg29}. The cosmological linear perturbations of $f(R,G)$ gravity minimally coupled with perfect fluid are analyzed for scalar-, vector- and tensor-type perturbations \cite{prd10}. If $f_{RR}f_{GG}\neq f_{RG}^2$, the scalar-type perturbations of $f(R,G)$ gravity will obey same dispersion relation as of non-relativistic de Broglie wave. The vector-type perturbations of $f(R,G)$ gravity may decay with time, in a same way as in General relativity. The tensor-type perturbations will be model  dependent and may be either superluminal or subnuminal \cite{prd10}.   \\
The higher-order theories of gravity possess high degree of non-linearity in the equations of motion which leads to a very difficult process for finding the analytical as well as numerical solutions. In this paper, we investigate the cosmic dynamics and properties of $f(R,G)$ gravity model by using dynamical system method. The linear perturbations around the fixed points are studied to explore the corresponding stability of fixed points. In particular, we proceed with two realistic forms of $f(R,G)$ function and compare their corresponding cosmic dynamics. The qualitative behavior and the global dynamics  of the cosmological model may be retrieved using dynamical system method. This method is well-utilized for the exploration of general cosmic dynamics in the cosmological models \cite{ds3,adf,scjpm,prd87,bbepjc,arxjls,a11,bm,1,48,ext6,int4,int5,a7,skdg,a10,a9,a8,a17,a14,a13,a18,a5,asijmpa,aschjp24,asgrg24,asskkcm,2020o1,2020o2,2020o3,2020o4,2020o5,2021o1,2022o1,2022o2,2022o3,2022o4,2022o5,2022o6,2023o1,2023o2,2023o3,2023o4,2023o5,2023o6,2023o7,2023o8,2024o1,asskascom}. This method provides a qualitative description of the universe evolution which is independent of the initial conditions \cite{b1,b2,sbrev}. Additionally, this method provides a simple way to obtain the numerical solutions. In the paper, we aim to identify different evolutionary phases of the universe possessed by the $f(R,G)$ gravity theory by considering two forms of $f(R,G)$ which are allowed by the Noether symmetry method \cite{3}. In particular, we aim to recast the gravitational field equations into autonomous system to identify the stability and corresponding cosmographic evolution. This will enable us to compare the results of $f(R,G)$ model with standard cosmological paradigm.

Different sections of the paper are arranged in the following way: In section (\ref{sec2}), we derive the field equations for non-flat Friedmann-Robertson-Walker (FRW) spacetime in $f(R,G)$ gravity. We utilize these equations to study the cosmological dynamical system of $f(R,G)=f_0R^\alpha G^\beta$ and  $f(R,G)=f_0R^n+f_1G^m$ models in section (\ref{sec3}) and (\ref{sec4}) respectively. We investigate for the full phase of possibilities for the universe evolution in these models in section (\ref{sec5}). We summarize the differences among these models with summary of the results in section (\ref{sec6}). 
\section{The $f(R,G)$ model and its dynamical system formulation}
\label{sec2}
In the present section, we provide details of the $f(R,G)$ gravity theory with general formalism of the gravitational equations of motion. The action for modified Gauss-Bonnet theory termed as the $f(R,G)$ gravity may be written as \cite{ee2,1,2}
\begin{equation}
	S=\frac{1}{2\kappa^2}\int d^4x\sqrt{-g}[f(R,G)+L_m]
	\label{eq1} 
\end{equation}
where $\kappa^2=8\pi G_n$ and $G_n,R,G$ denote the Newton's gravitational constant, the Ricci scalar and the Gauss-Bonnet invariant respectively. The quantity $\sqrt{-g}$ is determinant of the metric tensor $g_{ij}$, $f(R,G)$ denotes the general function of $R$ and $G$ along with $L_m$ denoting the matter Lagrangian density. We follow the units $8\pi G_n=c=k_b=\hbar=1$. The field equations of the $f(R,G)$ gravity may be written by varying the action (\ref{eq1}) with $g_{ij}$ as \cite{ee2}
\begin{eqnarray}
	G_{ij}=\kappa^2T_{ij}^m+\triangledown_i\triangledown_j f_R-g_{ij}\square f_R+2R\triangledown_i\triangledown_j f_G-2g_{ij}R\square f_G-4R_j^k\triangledown_k\triangledown_i f_G-4R_i^k\triangledown_k\triangledown_j f_G+ \nonumber \\ 4R_{ij}\square f_G+4g_{ij}R^{kl}\triangledown_k\triangledown_l f_G+4R_{iklj}\triangledown^k\triangledown^l f_G-\frac{1}{2}g_{ij}(Rf_R+Gf_G-f(R,G))+(1-f_R)G_{ij}
	\label{eq2}
\end{eqnarray}
where the energy-momentum tensor for the standard matter is given by \cite{1}
\begin{equation}
	T_{ij}^m=-\frac{2}{\sqrt{-g}}\frac{\delta (\sqrt{-g}L_m)}{\delta g^{ij}}
	\label{eq3}
\end{equation}    
and $G_{ij}$ is the Einstein's tensor given by $G_{ij}\equiv R_{ij}-\frac{1}{2}Rg_{ij}$. The quantities $f_R,f_G$ will denote the partial differentiation of function $f$ with $R$ and $G$ respectively. $\triangledown_i$ denotes the covariant derivative and $\square\equiv \triangledown_i\triangledown^i$.\\
The FRW metric is given by
\begin{equation}
	ds^2=-dt^2+a^2\left(\frac{dr^2}{1-kr^2}+r^2(d\theta^2+\sin^2\theta d\phi^2) \right)
	\label{eq4} 
\end{equation} 
where $a$ is the scale factor which depends on the cosmic time as a consequence of homogeneity property. The constant curvature parameter $k$ takes value $-1,0,+1$ for open, flat and closed spatial sections respectively. For (+2) signature FRW metric (\ref{eq4}), the Ricci scalar and Gauss-Bonnet invariant may be written as
\begin{equation}
	R=6\left(\dot{H}+2H^2+\frac{k}{a^2} \right), \qquad G=24(\dot{H}+H^2)\left(H^2+\frac{k}{a^2} \right)
	\label{eq5}  
\end{equation}
where overhead dot represents the differentiation with time. The expansion rate of the evolving universe is denoted by $H=\frac{\dot{a}}{a}$ and, $\frac{\ddot{a}}{a}=\dot{H}+H^2$. For metric (\ref{eq4}), the field equation (\ref{eq2}) may take the form
\begin{align}
	3f_R\left( H^2+\frac{k}{a^2}\right) & =\kappa^2\rho-3H\dot{f}_R+\frac{1}{2}\left( Rf_R+Gf_G-f\right) -12H\dot{f}_G\left(H^2+\frac{k}{a^2} \right) \label{eq6} \\
	f_R\left( 2\dot{H}+3H^2+\frac{k}{a^2}\right) & =-\kappa^2p -2H\dot{f}_R-\ddot{f}_R+\frac{1}{2}\left( Rf_R+Gf_G-f\right)-8H\dot{f}_G(\dot{H}+H^2) -\nonumber \\ & 4\ddot{f}_G\left(H^2+\frac{k}{a^2} \right) \label{eq7}
\end{align}
where $\rho$ and $p$ denote the energy density and pressure of the fluid. The stress-energy tensor of the fluid for (+2) signature metric is given by $T_{ij}=(\rho+p)u_iu_j+pg_{ij}$ with $u_iu^i=-1, u_0=1$ and $u_i=0,i=1,2,3$. For $k=0$, the field equations (\ref{eq6}) and (\ref{eq7}) will reduce into the $f(R,G)$ model field equations with flat FRW spacetime \cite{1,2,3}. The action (\ref{eq1}) reduces into the $f(G)$ gravity action for $R=0$. By substituting $R=0$ in the field equations (\ref{eq6} \& \ref{eq7}), one may write the field equations of $f(G)$ gravity (see Appendix I (\ref{apendix1})). The standard matter density may incorporate the matter (baryons and dark matter) and radiation with energy densities $\rho_m$ and $\rho_r$ respectively. The corresponding continuity equations may be written as
\begin{equation}
	\dot{\rho}_m+3H\rho_m=0, \qquad \dot{\rho}_r+4H\rho_r=0 
	\label{eq8}
\end{equation}
where $p_m=0$ and $p_r=\frac{\rho_r}{3}$ are the equation of states for the matter and radiation components respectively. By a redefinition of the fluid components, the field equations (\ref{eq6}) and (\ref{eq7}) may be rewritten as
\begin{align}
	3f_R\left( H^2+\frac{k}{a^2}\right) & = \kappa^2(\rho_m+\rho_r+\rho_{gb}) \label{eq9} \\
	2f_R\left( \dot{H}-\frac{k}{a^2}\right) & = -\kappa^2(\rho_m+\frac{4\rho_r}{3}+\rho_{gb}+p_{gb}) \label{eq10} 
\end{align}
with following definitions
\begin{align}
	\kappa^2\rho_{gb} & = \frac{1}{2}\left( Rf_R+Gf_G-f-6H\dot{f}_R -24H\dot{f}_G\left(H^2+\frac{k}{a^2} \right)\right),  \label{eq11} \\
	\kappa^2(\rho_{gb}+p_{gb}) & = -H\dot{f}_R+\ddot{f}_R-4H\dot{f}_G\left(-2\dot{H}+H^2+\frac{3k}{a^2} \right) +4\ddot{f}_G\left(H^2+\frac{3k}{a^2} \right) \label{eq12}
\end{align}
The importance of spatial curvature in the energy density and pressure of geometric dark energy may be observed from Eqs. (\ref{eq11}) and (\ref{eq12}). The equation of state parameter for the geometric dark energy may be defined as $\omega_{gb}=\frac{p_{gb}}{\rho_{gb}}$.\\
In order to investigate the qualitative behavior of the universe in $f(R,G)$ model with the FRW spacetime having $k=0$, we define the dynamical variables as
\begin{equation}
	\Omega_m=\frac{\kappa^2\rho_m}{3f_RH^2},\ \Omega_r=\frac{\kappa^2\rho_r}{3f_RH^2},\ x_1=\frac{\dot{f}_R}{f_RH},\ x_2=\frac{f}{6f_RH^2},\ x_3=\frac{R}{6H^2},\ x_4=\frac{Gf_G}{6f_RH^2},\ x_5=\frac{4H\dot{f}_G}{f_R} \label{eq13}
\end{equation}
With the variables (\ref{eq13}) and Eq. (\ref{eq6}) for flat universe, one may write a constraint equation as
\begin{equation}
	\Omega_m+\Omega_r-x_1-x_2+x_3+x_4-x_5=1
	\label{eq14}
\end{equation}
Above Eq. (\ref{eq14}) may also be written as $\Omega_m+\Omega_r+\Omega_{gb}=1$, where $\Omega_{gb}=-x_1-x_2+x_3+x_4-x_5$. In order to explore the cosmic dynamics, we define the deceleration parameter $(q)$ and effective equation of state parameter $(\omega)$ as
\begin{equation}
	q=-1+\frac{d}{dt}\frac{1}{H}, \qquad \omega=\frac{1}{3}(2q-1) \label{eq15}
\end{equation} 
respectively. The cosmographic parameters may also be written in the terms of dynamical variables (\ref{eq13}) using definitions of the Ricci scalar and Gauss-Bonnet invariant. In general, for a given form of $f(R,G)$ function, one may write $\dot{f}=f_R\dot{R}+f_G\dot{G}$, where $f_R$ and $f_G$ denote the partial derivatives of $f$ with respect to $R$ and $G$ respectively. Also, $\dot{f}_R=f_{RR}\dot{R}+f_{RG}\dot{G}$ and  $\dot{f}_G=f_{GR}\dot{R}+f_{GG}\dot{G}$. A careful application of differentiation rules are needed since $f(R,G)$ is a function of both $R$ and $G$.  \\
Using Eqs. (\ref{eq5}), (\ref{eq13}) and (\ref{eq15}), we may write the general expressions as
\begin{eqnarray}
	\frac{\dot{H}}{H^2}=x_3-2, \label{eq16}\\
	q=1-x_3, \label{eq17}\\
	\frac{G}{24H^4}=x_3-1, \label{eq18}\\
	\frac{\dot{G}}{GH}=-\frac{1}{q}\left( \frac{\dot{R}}{6H^3}-\frac{2\dot{H}}{H^2}\right) +\frac{2\dot{H}}{H^2}. \label{eq19}
\end{eqnarray}
Using redefined Eqs. (\ref{eq9}) and (\ref{eq10}) for $k=0$, we obtain
\begin{equation}
	\omega_{gb}=\frac{-1-\frac{\Omega _r}{3}-\frac{2\dot{H} }{3H^2}}{1-\Omega _m-\Omega _r}
	\label{eq20}
\end{equation}
For converting the dynamical variables (\ref{eq13}) into a cosmological dynamical system, we define the independent variable as $N\equiv \ln a$. As a consequence, $dN=Hdt$. This independent variable is suitable to study the expanding $(H>0)$ cosmologies. However, it is not suitable to study the bouncing models using dynamical system approach, since $H=0$ at the bounce \cite{a18,a17,a14,a11,a10,a9,a7,asijmpa,a16,a6,a5a}. In general, the dynamical system of $f(R,G)$ gravity model for variables (\ref{eq13}) may take the form
\begin{align}
	{\Omega_m}' & =-3\Omega_m-\Omega_mx_1-2\Omega_m\frac{\dot{H}}{H^2} \label{eq21} \\
	{\Omega_r}' & =-4\Omega_r-\Omega_rx_1-2\Omega_r\frac{\dot{H}}{H^2} \label{eq22} \\
	{x_1}' & =\frac{6\ddot{f}_R}{f}x_2-{x_1}^2-x_1\frac{\dot{H}}{H^2} \label{eq23} \\
	{x_2}' & =\frac{\dot{f}}{6f_RH^3}-x_1x_2-2x_2\frac{\dot{H}}{H^2} \label{eq24} \\
	{x_3}' & =\frac{\dot{R}}{6H^3}-2x_3\frac{\dot{H}}{H^2} \label{eq25} \\
	{x_4}' & =\frac{\dot{G}}{GH}x_4+\frac{G}{24H^4}x_5-x_1x_4-2x_4\frac{\dot{H}}{H^2} \label{eq26} \\
	{x_5}' & =\frac{\ddot{f}_G}{H\dot{f}_G}x_5-x_1x_5+x_5\frac{\dot{H}}{H^2} \label{eq27}
\end{align}
where prime denotes the derivative with respect to $\ln a$. Note that Eq. (\ref{eq27}) may be equivalently written as ${x_5}'=\frac{4\ddot{f}_G}{\dot{f}_R}-x_1x_5+x_5\frac{\dot{H}}{H^2}$.\\ The cosmological dynamical system of $f(R,G)$ gravity may be converted into the autonomous system after the specification of $f(R,G)$ function. By autonomous system, we mean a dynamical system where the independent variable does not appear explicitly. In particular, we proceed in the $f(R,G)$ model having form $f(R,G)=f_0R^\alpha G^\beta$ and $f(R,G)=f_0R^n+f_1G^m$ in section (\ref{sec3}) and (\ref{sec4}) respectively, where $f_0,f_1,n,m,\alpha$ and $\beta$ are the model parameters. 

\section{The dynamical system and cosmic dynamics of $f(R,G)=f_0R^\alpha G^\beta$ model}
\label{sec3}
For $f(R,G)=f_0R^\alpha G^\beta$ case, $f_{RR}f_{GG}\neq f_{RG}^2$ for $\alpha\neq 1,\beta\neq 1$. The general expressions for the quantities appearing in Eqs. (\ref{eq21}-\ref{eq27}) for $f(R,G)=f_0R^\alpha G^\beta$ may be written as
\begin{align}
	\frac{\dot{R}}{6H^3} & =\frac{x_1x_3}{\alpha-1}-\frac{\beta x_3}{\alpha-1}\frac{\dot{G}}{GH} \label{eq28} \\
	\frac{\dot{f}}{6f_RH^3} & =\frac{\dot{R}}{6H^3}+x_4\frac{\dot{G}}{GH} \label{eq29} \\
	\frac{\dot{G}}{GH} & =\frac{(1-\alpha)q}{\beta x_3+(1-\alpha)q}\left(\frac{x_1x_3}{(1-\alpha)q}+\frac{2\dot{H}}{H^2}\frac{q+1}{q} \right) \label{eq30} \\
	x_5 & = \frac{\beta (\beta-1)}{\alpha}\frac{\dot{G}}{GH}x_3\left(\frac{G}{24H^4} \right)^{-1} +\beta \left(\frac{G}{24H^4} \right)^{-1}\frac{\dot{R}}{6H^3} \label{eq31}
	\end{align}
In order to simplify the calculations, we proceed with $\beta=1-\alpha$ \cite{3} and by using Eqs. (\ref{eq16}-\ref{eq18}) in Eqs. (\ref{eq28}-\ref{eq31}), we may simplify as 
\begin{align}
	\frac{\dot{R}}{6H^3} & =-\frac{x_1 (x_3-1) x_3}{\alpha -1}-2 x_3 (x_3-2)^2 \label{eq32} \\
	\frac{\dot{f}}{6f_RH^3} & = -\frac{x_1 x_3 (x_3+x_4-1)}{\alpha -1}-2 (x_3-2)^2 (x_3+x_4)\label{eq33} \\
	\frac{\dot{G}}{GH} & = -\frac{x_1 x_3}{\alpha -1}-2 (x_3-2)^2 \label{eq34} \\
	x_5 & =\frac{x_1 x_3}{1-x_3} \label{eq35}
\end{align}
From above Eqs. (\ref{eq32}-\ref{eq35}), it is clear that the cosmological dynamical system (\ref{eq21}-\ref{eq27}) may become autonomous. For the given form of $f(R,G)$, the dynamical variable $x_5$ is dependent on other variables and therefore the system may be studied with the variables $(\Omega_m,\Omega_r,x_2,x_3,x_4)$. By using the constraint equation (\ref{eq14}) and Eq. (\ref{eq35}), we may write
\begin{align}
	x_1 & = -(x_3-1) \left(\Omega _m+\Omega _r-x_2+x_3+x_4-1\right) \label{eq36} \\
	x_5 & = x_3 \left(\Omega _m+\Omega _r-x_2+x_3+x_4-1\right) \label{eq37}
\end{align}
The system defined by Eqs. (\ref{eq21},\ref{eq22},\ref{eq24},\ref{eq25},\ref{eq26}) will possess 9 fixed points. These points with their corresponding existence criterion have been listed in Table (\ref{table1}).  
\begin{table}[h!]
	\begin{center}
		{\begin{tabular}{ccccccccc}
				\hline\noalign{\smallskip}
				Point & $\Omega_m$  & $\Omega_r$ & $x_1$ & $x_2$ &  $x_3$ & $x_4$ & $x_5$ & Existence\\
				\noalign{\smallskip}\hline\noalign{\smallskip}
				$P_1$ & $0$ & $0$ & $ 0$ & $\tau_1$ & $1$ & $\tau_1 -1$ & $-1$ & Always \\
				$P_2$ & $0$ & $0$ & $0$ & $\tau_2 +1$ & $2$ & $\tau_2$ & $0$ & Always  \\
				$P_3$ & $0$ & $0$ & $4$ & $-5$ & $0$ & $0$ & $0$ & Always \\
				$P_4$ & $0$ & $0$ & $-1$ & $0$ & $0$ & $0$ & $0$ & Always \\
				$P_5$ & $0$ & $1$ & $0$ & $0$ & $0$ & $0$ & $0$ & Always \\
				$P_6$ & $2$ & $0$ & $1$ & $0$ & $0$ & $0$ & $0$ & Always \\
				$P_7$ & $0$ & $5-\frac{4}{\alpha }$ & $-\frac{4}{\alpha -2}-4$ & $\frac{2 (\alpha -1)}{(\alpha -2) \alpha }$ & $\frac{2 (\alpha -1)}{\alpha -2}$ & $-\frac{2 (\alpha -1)^2}{(\alpha -2) \alpha }$ & $-\frac{4}{\alpha }+\frac{4}{\alpha -2}+8$ & $\alpha\neq 0,2$  \\
				$P_8$ & $\frac{7-8 \alpha }{1-2 \alpha }$ & $0$ & $-\frac{3}{\alpha -2}-3$ & $\frac{5-4 \alpha }{4 \alpha -2 \alpha ^2}$ & $\frac{5-4 \alpha }{4-2 \alpha }$ & $\frac{4 \alpha ^2-9 \alpha +5}{4 \alpha -2 \alpha ^2}$ & $\frac{3}{1-2 \alpha }+\frac{3}{\alpha -2}+6$ & $\alpha\neq 0,\frac{1}{2},2$ \\
				$P_9$ & $0$ & $0$ & $\frac{1}{1-2 \alpha }+1$ & $\frac{3-4 \alpha }{\alpha -2 \alpha ^2}$ & $\frac{3-4 \alpha }{1-2 \alpha }$ & $\frac{4 \alpha ^2-7 \alpha +3}{\alpha -2 \alpha ^2}$ & $\frac{1}{2 \alpha -1}-2$ & $\alpha\neq 0,\frac{1}{2},2$ \\
				
				\noalign{\smallskip}\hline
			\end{tabular}
			\caption{The value of dynamical variables at the fixed points of $f(R,G)=f_0R^\alpha G^{1-\alpha}$ model, where $\tau_1,\tau_2$ are some constants.}
			\label{table1}}      
	\end{center}
\end{table}
\subsection{Stability of the fixed points and the corresponding cosmic dynamics}
The existence of points restrict that $\alpha\neq 0,\frac{1}{2},2$. It simply mean that these points will not exist in the $f(R,G)$ gravity model given by $f(R,G)=f_0G^{-1}$, $f_0R^{\frac{1}{2}} G^{\frac{1}{2}}$ and $f_0R^{2} G^{-1}$. In addition, the cosmological characteristics of the fixed points of system are also depending on the parameter $\alpha$. The point-wise stability criterion with the cosmological implications have been listed below: 
\begin{enumerate}
	\item $P_1 $: For this point, the variables $x_2$ and $x_4$ are related to each other. In other words, there is some kind of interdependence between $f$ and $Gf_G$. The point represents the universe expanding with constant rate having $a\propto t$ and $H\propto \frac{1}{t}$. In other words, the effective EoS parameter highlights that $\omega=-\frac{1}{3}$ with the deceleration parameter lying on the transition line $q=0$. The eigenvalues of linearized matrix corresponding to this point are given by $\{0,-1,-2,2,2\}$. The point is saddle in nature in $\Omega_r-x_2-x_3-x_4$ plane.
	\item $P_2 $: This ever-existing point will have interdependence between $x_2$ and $x_4$. And, since at this point $R=12H^2$, the exponential expansion may be realized due to $\dot{H}=0$. The universe will expand and follows $q=-1$, influenced by the fluid behaving like cosmological constant in this $f(R,G)$ model. The eigenvalues of linearized matrix are $$\left\{0,-4,-3,-\frac{\sqrt{25 \alpha ^2-66 \alpha +41}+3 \alpha -3}{2 (\alpha -1)},\frac{\sqrt{25 \alpha ^2-66 \alpha +41}-3 \alpha +3}{2 (\alpha -1)}\right\}$$ The point may be stable in $\Omega_r-x_2-x_3-x_4$ plane if $\frac{41}{25}\leq \alpha <2$ and saddle otherwise. The attracting nature (for $\alpha\in [1.64,2)$) of this point may explain the present accelerating universe expansion in the model. 
	\item $P_3$: This ever-existing point yield the radiation-dominated universe in an effective sense, since $\omega=\frac{1}{3}$. At this point, $\Omega_r=0$ but the effective radiation dominated phase is the consequence of dominating Ricci scalar term. The eigenvalues are given by $\left\{-\frac{4 (\alpha -2)}{\alpha -1},-8,-5,-4,-3\right\}$. The point is saddle for $1<\alpha <2$ and stable otherwise. The point may be an attractor of the model for $\alpha\in (-\infty,1)\cup (2,\infty)$. Therefore, there is a possibility of late-time decelerating universe subjected to the proper choice of initial conditions.
	\item $P_4 $: The eigenvalues  of the Jacobian matrix are $\left\{\frac{3-4 \alpha }{\alpha -1},-3,1,2,5\right\}$. The point is saddle in nature and this behavior is independent of the model parameter value. The point may correspond to the radiation-dominated decelerating universe, since $\omega=\frac{1}{3}$ in an effective sense. The role of $x_1$ is important for the effective radiation-dominated era governed by this point, since $\omega_{gb}=\frac{1}{3}$. 
	\item $P_5$: This ever-exiting point has eigenvalues $\{-4,-4,4,-1,1\}$, which are composed of positive and negative signs. It means that the point will be a saddle point. The stability nature of this point is not affected by the model parameter $\alpha$. At this point, $\Omega_r=1$ and effective EoS parameter $\omega=\frac{1}{3}$ highlight that it will represent the radiation dominated, decelerating phase  of the universe evolution. The universe will expand with $a\propto t^{\frac{1}{2}}$. In this phase, the role of terms due to $R$ and $G$ are not important.
	\item $P_6 $: This point will exist for the all values of $\alpha$. The eigenvalues $\left\{\frac{5-4 \alpha }{\alpha -1},-1,-5,-2,3\right\}$ show that the point is saddle in nature. It will correspond to the decelerating universe expansion, which is influenced by the effective radiation-like fluid. At this point, $\Omega_m=2$. Here, it is important to stress that $\Omega_m=\frac{\kappa^2\rho_m}{3f_RH^2}$, which is different from the standard matter density parameter $\Omega_{m'}=\frac{\kappa^2\rho_m}{3H^2}$. It means that the existence of standard matter dominated phase is affected due to the contribution of factor $\frac{\partial f}{\partial R}$.
	\item $P_7 $: The point will exist if $\alpha\neq 0,2$. The eigenvalues are given by $$\left\{1,4,\frac{4 \alpha }{\alpha -2},\frac{\alpha -\sqrt{\alpha } \sqrt{81 \alpha -64}}{2 (\alpha -2)},\frac{\alpha +\sqrt{\alpha } \sqrt{81 \alpha -64}}{2 (\alpha -2)}\right\}$$ 
	The point is saddle in nature. The cosmic dynamics corresponding to this point depends on $\alpha$ and it may correspond to matter-dominated universe for $\alpha=\frac{2}{3}$. However, the point will never correspond to the de Sitter universe having $q=-1$ and $\omega=-1$ in an effective sense. \\
	Among other possibilities, the effective quintessence (phantom) energy dominated evolution of universe may constrain $\alpha<0$ ($\alpha>2$) respectively. 
	\item $P_8 $: The point will exist for $\alpha\neq 0,\frac{1}{2},2$. The universe evolution may scale according to the model parameter value $\alpha$. And, for $\alpha=1$ (and correspondingly $f(R,G)=f_0R$), the point would correspond to the matter-dominated universe. In other words, this point would yield matter dominated phase of General relativity, since $\alpha=1$ in $f(R,G)=f_0R^\alpha G^{1-\alpha}$ would yield the Lagrangian of General relativity. The effective quintessence (phantom) energy dominated universe evolution will constrain  $\alpha<\frac{1}{2}$ ($\alpha>2$) respectively. \\ The eigenvalues are given by $\left\{-1,3,\frac{3 \alpha }{\alpha -2},\frac{A+\sqrt{B}+12 \alpha }{C},\frac{A+\sqrt{B}-12 \alpha }{C}\right\}$
	where $A=12 \alpha ^6-72 \alpha ^5+159 \alpha ^4-159 \alpha ^3+72 \alpha ^2$, $B=\alpha ^2 \left(2 \alpha ^2-5 \alpha +2\right)^4 \left(256 \alpha ^4-928 \alpha ^3+1233 \alpha ^2-710 \alpha +149\right)$, $C=4 (1-2 \alpha )^2 (\alpha -2)^3 (\alpha -1) \alpha $. The point is saddle in nature. 
	\item $P_9 $: The point will exist for $\alpha\neq 0,\frac{1}{2},2$. The eigenvalues are given by $$\left\lbrace -\frac{2 (\alpha -2)}{2 \alpha -1},\frac{2 \alpha }{1-2 \alpha },\frac{6-8 \alpha }{2 \alpha -1},\frac{8-10 \alpha }{2 \alpha -1},\frac{7-8 \alpha }{2 \alpha -1}\right\rbrace $$
	The point is stable for either $\alpha <0 $ or $\alpha >2$ and saddle for $0<\alpha<2$. The matter-dominated universe evolution may correspond to $\alpha = \frac{5}{6}$. Among other possibilities, the point would never cosmologically correspond to the $\Lambda$CDM universe. The quintessence (phantom) dominated universe may be realized for $\alpha>1$ ($\alpha<\frac{1}{2}$) respectively. 
\end{enumerate}
\begin{table}[h!]
	\begin{center}
		{\begin{tabular}{cccccccc}
		\hline\noalign{\smallskip}
		Point & $q$  & $\omega$ & $\omega_{gb}$ &  $a$ & $H$ & $r$ & $s$ \\
		\noalign{\smallskip}\hline\noalign{\smallskip}
		$P_1$ & $0$  & $-\frac{1}{3}$ & $-\frac{1}{3}$ & $t$ & $\frac{1}{t}$ & $0$ & $\frac{2}{3}$  \\
		$P_2$ & $-1$  & $-1$ & $-1$ & $e^{H_0t}$ & $H_0$ & $1$ & $0$  \\
		$P_3$ & $1$ & $\frac{1}{3}$ & $\frac{1}{3}$ & $t^{\frac{1}{2}}$ & $\frac{1}{2t}$ & $3$ & $\frac{4}{3}$ \\
		$P_4$ & $1$ & $\frac{1}{3}$ & $\frac{1}{3}$ & $t^{\frac{1}{2}}$ & $\frac{1}{2t}$ & $3$ & $\frac{4}{3}$ \\
		$P_5$ & $1$ & $\frac{1}{3}$ & $-$ & $t^{\frac{1}{2}}$ & $\frac{1}{2t}$ & $3$ & $\frac{4}{3}$ \\
		$P_6$ & $1$ & $\frac{1}{3}$ & $-\frac{1}{3}$ & $t^{\frac{1}{2}}$ & $\frac{1}{2t}$ & $3$ & $\frac{4}{3}$ \\
		$P_7$ & $\frac{\alpha }{2-\alpha }$ & $\frac{4}{6-3 \alpha }-1$ & $\frac{2 (\alpha -1)}{3 (\alpha -2)}$ & $t^{\frac{2-\alpha}{2}}$ & $\frac{2-\alpha}{2t}$ & $\frac{\alpha  (\alpha +2)}{(\alpha -2)^2}$ & $\frac{4}{6-3 \alpha }$ \\
		$P_8$ & $\frac{3}{4-2 \alpha }-1$ & $\frac{1}{2-\alpha }-1$ & $\frac{1}{2 (\alpha -2)}+\frac{1}{3}$ & $t^{\frac{4-2 \alpha }{3}}$ & $\frac{4-2 \alpha }{3}$ & $\frac{(\alpha +1) (2 \alpha -1)}{2 (\alpha -2)^2}$ & $\frac{1}{2-\alpha }$ \\
		$P_9$ & $\frac{1}{2 \alpha -1}-1$ & $\frac{5-6 \alpha }{6 \alpha -3}$ & $\frac{5-6 \alpha }{6 \alpha -3}$ & $t^{2 \alpha -1}$ & $\frac{{2 \alpha -1}}{t}$ & $\frac{2 (\alpha -1) (2 \alpha -3)}{(1-2 \alpha )^2}$ & $\frac{2}{6 \alpha -3}$ \\
		
		\noalign{\smallskip}\hline
		\end{tabular}
		\caption{The cosmological quantities at the fixed points of $f(R,G)=f_0R^{\alpha}G^{1-\alpha}$ model.}
		\label{table2}}      
		\end{center}
\end{table}
The cosmological quantities and parameters evaluated at the fixed point have been summarized in Table (\ref{table2})
\section{The dynamical system and cosmic dynamics of $f(R,G)=f_0R^n+f_1G^m$ model}
\label{sec4}
For $f(R,G)=f_0R^n+f_1G^m$ case, $f_{RR}f_{GG}\neq 0$. The general form of the quantities appearing in Eqs. (\ref{eq21}-\ref{eq27}) for $f(R,G)=f_0R^n+f_1 G^m$ model are 
\begin{align}
	\frac{\dot{R}}{6H^3} & =\frac{x_1x_3}{n-1} \label{eq38} \\
	\frac{\dot{G}}{GH} & =2 (x_3-2)-\frac{x_3 \left(\frac{x_1}{n-1}-2\right)+4}{1-x_3} \label{eq40} \\
	\frac{\dot{f}}{6f_RH^3} & =\frac{\dot{R}}{6H^3}+x_4\frac{\dot{G}}{GH} \label{eq39}  
\end{align}
For simplicity of the analysis ahead, we proceed with $m=1$ and by using Eqs. (\ref{eq38},\ref{eq40}) in Eq. (\ref{eq39}), we may write 
\begin{align}
	\frac{\dot{f}}{6f_RH^3} & = \frac{2 (n-1) (x_3-2)^2 x_4+x_1 x_3 (x_3+x_4-1)}{(n-1) (x_3-1)}\label{eq41}  
\end{align}
For this form of $f(R,G)$, the dynamical variable $x_5$ is independent with other variables and therefore the system may be studied with the variables $(\Omega_m,\Omega_r,x_2,x_3,x_4,x_5)$. For this model, the constraint equation is given by Eq. (\ref{eq14}).\\
The system (\ref{eq21},\ref{eq22},\ref{eq24},\ref{eq25},\ref{eq26},\ref{eq27}) when solved by using Eqs. (\ref{eq38},\ref{eq40},\ref{eq41}) will yield 9 fixed points. These points with their corresponding existence criterion have been listed in Table (\ref{table3}).  
\begin{table}[h!]
	\begin{center}
		{\begin{tabular}{ccccccccc}
				\hline\noalign{\smallskip}
				Point & $\Omega_m$  & $\Omega_r$ & $x_1$ & $x_2$ &  $x_3$ & $x_4$ & $x_5$ & Existence\\
				\noalign{\smallskip}\hline\noalign{\smallskip}
				$Q_1$ & $0$ & $0$ & $0$ & $\tau +1$ & $2$ & $\tau$ & $0$ &  Always \\
				$Q_2$ & $0$ & $0$ & $-2$ & $-\frac{4}{5}$ & $0$ & $-\frac{3}{5}$ & $\frac{6}{5}$ & Always  \\
				$Q_3$ & $0$ & $0$ & $4$ & $-5$ & $0$ & $0$ & $0$ & Always \\
				$Q_4$ & $0$ & $0$ & $-1$ & $0$ & $0$ & $0$ & $0$ & Always \\
				$Q_5$ & $0$ & $1$ & $0$ & $0$ & $0$ & $0$ & $0$ & Always \\
				$Q_6$ & $2$ & $0$ & $1$ & $0$ & $0$ & $0$ & $0$ & Always \\
				$Q_7$ & $0$ & $-\frac{2}{n^2}+\frac{8}{n}-5$ & $\frac{4}{n}-4$ & $\frac{2 (n-1)}{n^2}$ & $2-\frac{2}{n}$ & $0$ & $0$ & $n\neq 0$  \\
				$Q_8$ & $-\frac{8 n^2-13 n+3}{2 n^2}$ & $0$ & $\frac{3}{n}-3$ & $\frac{4 n-3}{2 n^2}$ & $2-\frac{3}{2 n}$ & $0$ & $0$ & $n\neq 0$ \\
				$Q_9$ & $0$ & $0$ & $\frac{3}{1-2 n}+1$ & $\frac{4 n-5}{2 n^2-3 n+1}$ & $\frac{n (4 n-5)}{2 n^2-3 n+1}$ & $0$ & $0$ & $n\neq 0,\frac{1}{2},1$ \\
				
				\noalign{\smallskip}\hline
			\end{tabular}
			\caption{The value of dynamical variables at the fixed points of $f(R,G)=f_0R^n+f_1G$ model.}
			\label{table3}}      
	\end{center}
\end{table}
\subsection{Stability of fixed points and the corresponding cosmic dynamics}
The existence of points restrict that $n\neq 0,\frac{1}{2},1$. The fixed points and their stability criterion with corresponding cosmic dynamics are as follows:
\begin{enumerate}
	\item $Q_1$: This point will have inter-dependent $x_2$ and $x_4$ terms, which means that the components $f$ and $Gf_G$ are related to each other. This point will correspond to $R=12H^2$, which will yield $\dot{H}=0$ meaning that this may represent the accelerating universe expansion under the influence of effective cosmological constant-like fluid. In other words, the point may yield $\Lambda$CDM universe evolution. The eigenvalues are given by $$\left\{0,0,-4,-3,-\frac{\sqrt{25 n^2-66 n+41}+3 n-3}{2 (n-1)},\frac{\sqrt{25 n^2-66 n+41}-3 n+3}{2 (n-1)}\right\}$$ The point is stable in $x_2-x_3-x_4-x_5$ plane for $\frac{41}{25}\leq n<2$ and saddle otherwise. 
	\item $Q_2 $: This ever-existing point will correspond to the effective radiation dominated universe, since $\omega=\frac{1}{3}$. This phase is not the standard radiation phase, since $\Omega_r=0$. The eigenvalues are given by $\left\{-2,1,2,3,6,\frac{4 n-6}{n-1}\right\}$. The point is a saddle point.
	\item $Q_3$: The point will exist for all $n$. The eigenvalues are $\left\{-8,-6,-5,-4,-3,\frac{4 n}{n-1}\right\}$. It is stable for $0<n <1$ and saddle otherwise, and may represent a radiation-dominated universe where $a\propto t^{\frac{1}{2}}$.
	\item $Q_4 $: This always existing point will have the eigenvalues $\left\{-1,-3,1,2,5,\frac{4 n-5}{n-1}\right\}$, which means that the point is saddle in nature. The universe will have effective fluid following $\omega=\frac{1}{3}$ and decelerating with $q=1$.
	\item $Q_5$: The point will exist for all $n$ and the stability nature is independent of parameter $n$. The radiation component variable $\Omega_r=1$, yielding a radiation dominated, decelerating universe. The universe will expand with $a\propto t^{\frac{1}{2}}$ and $H=\frac{1}{2t}$. The eigenvalues of Jacobian matrix at $Q_5$ are given by $\{-4,4,4,-2,-1,1\}$. The point is saddle in nature. 
	\item $Q_6 $: The eigenvalues are given by $\left\{-1,-5,-3,-2,3,\frac{4 n-3}{n-1}\right\}$. The point is saddle in nature. The point correspond to universe scaling with $a\propto t^{\frac{1}{2}}$ having radiation dominated, decelerating universe but $\Omega_r=0$ and $\Omega_m=2$.  
	\item $Q_7 $: The condition for existence of this point implies that $n\neq 0$. The eigenvalues are $$\left\{4-\frac{8}{n},1,4,4-\frac{6}{n},\frac{n^4-2 n^3-\sqrt{3} A}{2 n^4},\frac{n^4-2 n^3+\sqrt{3} A}{2 n^4}\right\}$$ where $A=\sqrt{n^6 \left(27 n^2-44 n+12\right)}$. This point will be saddle in nature.\\
	The point may yield matter dominated universe for $n=\frac{4}{3}$. For $n<0$, the phantom universe evolution may be realized and for quintessence dominated universe $n>2$. The $\Lambda$CDM universe expansion will never be realized corresponding to this point.
	\item $Q_8 $: The existence criterion constrains that $n\neq 0$. The eigenvalues are $$\left\{3-\frac{6}{n},-1,3-\frac{9}{2 n},3,-\frac{\sqrt{B}+3 n-3}{4 (n-1) n},\frac{\sqrt{B}-3 n+3}{4 (n-1) n}\right\}$$
	where $B=256 n^4-864 n^3+1025 n^2-498 n+81$. It is a saddle point.\\
	The point may correspond to the effective matter dominated decelerating evolution for $n=1$. The quintessence (phantom) dominated universe may be realized for $n>\frac{3}{2}$ ($n<0$) respectively. The point will never represent an universe expansion having $q=-1$.
	\item $Q_9 $: The point will exist for $n\neq 0,\frac{1}{2},1$. The eigenvalues are given by $$\left\{\frac{-2 (n-2)^2}{(n-1) (2 n-1)},\frac{-2 n^2+7 n-6}{2 n^2-3 n+1},\frac{-2 \left(5 n^2-8 n+2\right)}{2 n^2-3 n+1},\frac{-8 n^2+13 n-3}{2 n^2-3 n+1},\frac{5-4 n}{n-1},\frac{-2 (n-2) n}{(n-1) (2 n-1)}\right\} $$
		The point is stable for either $n <0 $ or $n >2$ and saddle for $0<n <2 $. The matter dominated universe expansion may be realized either for $n=\frac{1}{12} \left(7-\sqrt{73}\right)$ or $n=\frac{1}{12} \left(\sqrt{73}+7\right)$. The quintessence dominated universe evolution may yield either $n<\frac{1}{2} \left(1-\sqrt{3}\right)$ or $\frac{1}{2} \left(\sqrt{3}+1\right)<n<2$. The phantom kind of dark energy will exist in the universe corresponding to this point for either $\frac{1}{2}<n<1$ or $n>2$.
\end{enumerate}
The cosmological parameters and the behavior of scale factor at the fixed points have been summarized in Table (\ref{table4}).
\begin{table}[h!]
	\begin{center}
		{\begin{tabular}{cccccccc}
				\hline\noalign{\smallskip}
				Point & $q$  & $\omega$ & $\omega_{gb}$ &  $a$ & $H$ & $r$ & $s$ \\
				\noalign{\smallskip}\hline\noalign{\smallskip}
				$Q_1$ & $-1$  & $-1$ & $-1$ & $e^{H_0t}$ & $H_0$ & $1$ & $0$  \\  
				$Q_2$ & $1$ & $\frac{1}{3}$ & $\frac{1}{3}$ & $t^{\frac{1}{2}}$ & $\frac{1}{2t}$ & $3$ & $\frac{4}{3}$ \\
				$Q_3$ & $1$ & $\frac{1}{3}$ & $\frac{1}{3}$ & $t^{\frac{1}{2}}$ & $\frac{1}{2t}$ & $3$ & $\frac{4}{3}$ \\
				$Q_4$ & $1$ & $\frac{1}{3}$ & $\frac{1}{3}$ & $t^{\frac{1}{2}}$ & $\frac{1}{2t}$ & $3$ & $\frac{4}{3}$ \\
				$Q_5$ & $1$ & $\frac{1}{3}$ & $-$ & $t^{\frac{1}{2}}$ & $\frac{1}{2t}$ & $3$ & $\frac{4}{3}$ \\
				$Q_6$ & $1$ & $\frac{1}{3}$ & $-\frac{1}{3}$ & $t^{\frac{1}{2}}$ & $\frac{1}{2t}$ & $3$ & $\frac{4}{3}$ \\
				$Q_7$ & $\frac{2}{n}-1$ & $\frac{4}{3 n}-1$ & $\frac{n-1}{9 n-3}$ & $t^{\frac{n}{2}}$ & $\frac{n}{2t}$ & $\frac{(n-4) (n-2)}{n^2}$ & $\frac{4}{3 n}$ \\
				$Q_8$ & $\frac{3}{2 n}-1$ & $\frac{1}{n}-1$ & $\frac{2 n}{3-10 n}$ & $t^{\frac{2 n}{3}}$ & $\frac{2n}{3}$ & $\frac{(n-3) (2 n-3)}{2 n^2}$ & $\frac{1}{n}$ \\
				$Q_9$ & $\frac{3}{1-2 n}+\frac{1}{n-1}-1$ & $\frac{n (7-6 n)+1}{6 n^2-9 n+3}$ & $\frac{n (7-6 n)+1}{6 n^2-9 n+3}$ & $t^{\frac{2 n^2-3 n+1}{2-n}}$ & $\frac{2 n^2-3 n+1}{(2-n)t}$ & $\frac{(n+1) (2 n-3)}{\left(2 n^2-3 n+1\right)}\times $ & $\frac{4-2 n}{6 n^2-9 n+3}$ \\
				 &  &  &  &  &  & $ \frac{(2 (n-1) n-1)}{\left(2 n^2-3 n+1\right)}$ &  \\
				
				\noalign{\smallskip}\hline
			\end{tabular}
			\caption{The cosmological quantities at the fixed points of $f(R,G)=f_0R^n+f_1G$ model.}
			\label{table4}}      
	\end{center}
\end{table}

\section{General issues}
\label{sec5}
\subsection{Cosmography}
The cosmic evolution in a model may be studied by using the cosmographic parameters such as the Hubble $(H)$, deceleration $(q)$, jerk $(j)$ parameters. These cosmological parameters may be defined as \cite{arle24,a15,a12,a10a}
\begin{equation}
	q=-\frac{\ddot{a}}{aH^2}, \quad j=\frac{\dddot{a}}{aH^3}
	\label{eqgi1}
\end{equation} 
where dot is representing the time derivative. The accelerating (decelerating) universe expansion may be characterized by $q<0 \ (q>0)$ respectively. The de Sitter expansion follows $q=-1$. The $\Lambda$CDM dynamics at late-times is governed by $q=-1$, $j=1$. \\
In the present cosmological dynamical system, the time derivative is related to $N\equiv \ln a$ implying $dN=Hdt$ and thus by using Eqs. (\ref{eq16}) and (\ref{eq17}), we one identify the deceleration-acceleration scenarios in the model. The jerk parameters defined above may be written as
\begin{equation}
	j=-\frac{dq}{dN}+q(1+2q)
	\label{eqgi2}
\end{equation} 
The summary of jerk parameter values evaluated at fixed points have been summarized in Table (\ref{table2}) and (\ref{table4}) respectively. 
\subsection{Statefinder diagnostic}
The statefinder diagnostic parameters $\{r,s\}$ \cite{al2003} are constructed using the geometrical quantities and are used to characterize the dark energy nature in the model. These diagnostic parameters classify among the quintessence and Chaplygin gas models. The quintessence kind of dark energy possess the equation of state parameter lying in range $\left(-1,-\frac{1}{3} \right)$. The Chaplygin gas unifies the decelerating phase to the accelerating phase of the universe in models. The $\Lambda$CDM dynamics corresponds to $r=1,s=0$. Note that the mathematical definitions of the statefinder $r$ is same as that of jerk parameter. The statefinder $s$ may be defined as 
\begin{equation}
	s=\frac{r-1}{3\left( q-\frac{1}{2}\right) } 
	\label{eqgi3}
\end{equation} 
We utilize Eqs. (\ref{eqgi2},\ref{eqgi3}) to evaluate these quantities at the fixed points. The summary of these results are given in Table (\ref{table2}) and (\ref{table4}) for model I and II respectively. 
\subsection{Scale factor evolution}
The definition of deceleration parameter (\ref{eq17}) may be used to identify the power-law and exponential law evolution in the $f(R,G)$ models at the fixed points. The $f(R,G)$ models studied are possessing the dark energy dominated phases along with the matter and radiation phases. The corresponding scale factor and Hubble parameter evolution are summarized in Table (\ref{table2}) and (\ref{table4}) for model I and II respectively.
 
\section{Conclusions}
\label{sec6}
In the present paper, we studied the cosmological implications of the modified gravity consisting of the Ricci scalar and Gauss invariant term. In the theory termed as $f(R,G)$ gravity, we derived the field equations for the non-flat FRW spacetime model. Due to non-linear nature of the field equations, we adopted the dynamical system technique to study the class of phenomenological functions of $R$ and $G$ in the $f(R,G)$ gravity model. In particular, the linear stability theory is used to study the fixed points and their stability character in the autonomous dynamical systems of $f(R,G)$ models in flat-FRW spacetime. For this, we assumed the forms $f(R,G)=f_0R^\alpha G^{1-\alpha}$ and $f(R,G)=f_0R^n+f_1G$ motivated from the Noether symmetry analysis to study the qualitative evolution of the universe by searching for the realizations of late-times dark energy era with intermediate radiation and matter eras.  \\
In the $f(R,G)=f_0R^\alpha G^{1-\alpha}$ model, the existence of fixed points $P_i,i=1,...,9$ restricts that $\alpha\neq 0,\frac{1}{2},2$. The model may be studied by the five independent variables in the autonomous form which may yield a late-time attractor corresponding to the effective cosmological constant-like fluid in the model. This formalism may yield the fixed point corresponding to the radiation dominated phase. The model may not yield the standard matter dominated phase, unless $\alpha=1$ and this condition will reduce the $f(R,G)$ model into the General relativity model. In other words, the non-linearity associated in the model due to $R$ and $G$ does not accommodate the fixed point corresponding to the standard matter dominated phase in the $f(R,G)$ model governed by this class of functions. This is an interesting aspect of non-linearity in $f(R,G)$ gravity.

In the $f(R,G)=f_0R^n+f_1G^m$ model, we convert the dynamical system into autonomous form by $m=1$ assumption, since the non-linearity associated with the Gauss-Bonnet term for $m\neq 1$ does not allow the dynamical system to be autonomous. The dynamical system in the autonomous form possess $9$ fixed points $Q_i, \ i=1,...,9$. The existence criterion of these points suggest that $n\neq 0,\frac{1}{2},1$. In the model, there will exist a point corresponding to the late-time accelerated universe expansion. However, the fixed point is stable either for $n<0$ or $n>2$. For $n<0$, the dark energy will be of quintessence nature and for $n>2$, the dark energy will be of phantom nature. The $f(R,G)$ gravity model governed by this class of function may explain the intermediate matter and radiation dominated era of universe evolution. This is an important difference of $f(R,G)=f_0R^n+f_1G$ model as compared to the $f(R,G)=f_0R^\alpha G^{1-\alpha}$ model.

\section*{Acknowledgments} 
The author would like to thank the IUCAA, Pune for support under Visiting associateship program. Author also acknowledge the facilities of ICARD, GLA University supported by IUCAA, Pune. 
\section*{Declaration of interests}
There are no competing financial interests to declare.
\section*{Data availability statement}
This theoretical study is conducted without using any data sets.
\section*{Appendix I}
\label{apendix1}
The field equations in $f(G)$ gravity for non-flat FRW spacetime may take the form
\begin{align}
	3 H^2+\frac{3k}{a^2} & =\kappa^2\rho+\frac{1}{2}\left( Gf_G-f\right) -12H\dot{f}_G\left(H^2+\frac{k}{a^2} \right) \label{eqA1} \\
	 2\dot{H}+3H^2+\frac{k}{a^2} & =-\kappa^2p +\frac{1}{2}\left( Gf_G-f\right)-8H\dot{f}_G(\dot{H}+H^2) - 4\ddot{f}_G\left(H^2+\frac{k}{a^2} \right) \label{eqA2}
\end{align}
where $\rho$ and $p$ denote the energy density and pressure of the fluid.

\end{document}